# Superconductivity with the enhanced upper critical field in the Pt-Doping CuRh$_2$Se$_4$ spinel

*Yiyi He[1,#], Yi-Xin You[2,#], Lingyong Zeng[1], Shu Guo[3,4], Huawei Zhou[1], Kuan Li[1], Yanhao Huang[1], Peifeng Yu[1], Chao Zhang[1], Chao Cao[2,5], Huixia Luo[1*]*

[1]*School of Materials Science and Engineering, State Key Laboratory of Optoelectronic Materials and Technologies, Key Lab of Polymer Composite & Functional Materials, Guangzhou Key Laboratory of Flexible Electronic Materials and Wearable Devices, Sun Yat-Sen University, No. 135, Xingang Xi Road, Guangzhou, 510275, P. R. China.*
[2]*School of Physics and Hangzhou Key Laboratory of Quantum Matters, Hangzhou Normal University, Hangzhou, 311121, P. R. China*
[3]*Shenzhen Institute for Quantum Science and Engineering (SIQSE), Southern University of Science and Technology, Shenzhen, P. R. China*
[4]*International Quantum Academy (SIQA), and Shenzhen Branch, Hefei National Laboratory, Futian District, Shenzhen, P. R. China*
[5]*Department of Physics and Center for Correlated Matter, Zhejiang University, Hangzhou, 310013, P. R. China*

[#] These authors contributed equally to this work.

[*]*Corresponding author/authors complete details (Telephone; E-mail:) (+86)-2039386124; E-mail address: luohx7@mail.sysu.edu.cn*

**Abstract**

We report the effect of Pt doping on the superconductivity in $CuRh_2Se_4$ spinel using a combined experimental and theoretical study. Our XRD results reveal that the $Cu(Rh_{1-x}Pt_x)_2Se_4$ ($0 \leq x \leq 0.35$) crystallizes in the structure with a space group of $Fd\bar{3}m$ (No. 227), and the lattice parameter $a$ increases with Pt doping. The resistivity and magnetic susceptibility measurement results verify that the superconducting transition temperature ($T_c$) forms a dome-like shape with a maximum value of 3.84 K at $x = 0.06$. It is also observed that the Pt-doping slightly reduces the lower critical magnetic field from 220 Oe in $CuRh_2Se_4$ to 168 Oe in $Cu(Rh_{0.94}Pt_{0.06})_2Se_4$, while it significantly enhances the upper critical magnetic field, reaching the maximum of 4.93 T in the $Cu(Rh_{0.94}Pt_{0.06})_2Se_4$ sample. The heat capacity result indicates that the sample $Cu(Rh_{0.91}Pt_{0.09})_2Se_4$ is a bulk superconductor. First-principles calculations suggest that the Pt-doping leads to a red-shift of a density of state peak near the Fermi level, consistent with the dome-like $T_c$ observed experimentally.

# I. Introduction

As a family of well-studied materials, spinel structural compounds are well-known for their excellent electrical, magnetic, thermal, and unique properties such as colossal magnetostriction, gigantic Kerr rotation, and multiferroic [1-6]. The chemical composition of spinel could be described by $AB_2X_4$, where metal ions occupy $A$ and $B$ sites, and $X$ sites are filled with elements such as oxygen, sulfur, selenium, tellurium. Even though thousands of spinels have been discovered, only a few spinels exhibit superconductivity (SC). So far, $Li_{1+x}Ti_{2-x}O_{4-\delta}$ is the only bulk superconducting oxide spinel and hosts the highest superconducting critical transition temperature ($T_c$) of 13.7 K among the spinel compound family [7]. Recently, SC is observed in $MgTi_2O_4$ oxide spinel film grown on the $MgAl_2O_4$ substrate, which can be achieved by engineering a superlattice of $MgTi_2O_4$ and $SrTiO_3$ [8]. The other spinel superconductors that have been reported so far are undoped ternary sulpo- and selenospinels $CuRh_2S_4$, $CuRh_2Se_4$ and $CuCo_2S_4$, as well as electron-doped $CuIr_2S_4$ and $CuIr_2Se_4$ [9-18].

Significantly, the heavy metal chalcogenide spinel $CuIr_2S_4$ has attracted considerable interest due to the metal-insulator (M-I) transition ($T_{MI}$) at T ≈ 230 K, accompanied by an intricate structural transition that simultaneously engenders both charge ordering and metal-metal pairing [19,20]. Besides, the peculiar $T_{MI}$ in $CuIr_2S_4$ increases with pressure and disappears gradually at higher pressure [21], whereas the $T_{MI}$ is suppressed by Zn substitution for Cu in the $Cu_{1-x}Zn_xIr_2S_4$ series, and SC is induced with a maximum $T_c$ of 3.4 K near $x = 0.3$ [16]. It is reminiscent of the emergence of SC in the doped chalcogenide series [17] and the high-temperature (HT) superconductors (e.g., cuprates and iron-based pnictides) [22,23], where the charge density wave (CDW) or the magnetism was suppressed by different dopants as a result of doping-induced destabilization of the charge-ordered or spin-paired state.

In contrast, such M-I transition is absent in the isostructural selenospinel $CuRh_2Se_4$, in which $T_c$ is around 3.5 K [24]. In addition, $T_c$ in $CuRh_2Se_4$ can be enhanced to 4.9 K at 6.5 GPa and possible superconductor–insulator transition emerges upon further compression with a key characteristic of broad peak in resistivity [25], which implies that $CuRh_2Se_4$ may be in close proximity to some kind of instabilities owing to an external disturbance despite the absence of charge-ordering or metal-metal pairing in $CuRh_2Se_4$. However, there are rare reports on the effect of chemical doping on the physical properties of $CuRh_2Se_4$ so far. Thus, this situation inspires us to explore the chemical doping effect on the physical properties of $CuRh_2Se_4$. In addition, compared with $CuRh_2S_4$, the Rh-4$d$ and Cu-3$d$ orbitals in $CuRh_2Se_4$ contribute more to the electronic states near the Fermi level, as proposed by previous first-principles calculations [26]. Besides, as far as we know, the 3$d$, 4$d$, 5$d$ transition metal dopants (e.g., Co, Ni, Ru, Ir) play an essential role in tuning SC in the iron-based superconductors [27-31]. Therein, Pt substitution for Fe in $BaFe_2As_2$ is reported to cause a significant increment on the $T_c$ and the upper critical magnetic field ($\mu_0H_{c2}$) [32]. Therefore, it will be interesting to study the properties of $Cu(Rh_{1-x}Pt_x)_2Se_4$ chalcogenides.

In this study, we synthesize $Cu(Rh_{1-x}Pt_x)_2Se_4$ ($0 \leq x \leq 0.35$) by a conventional solid-state reaction method. The crystal structure, resistivity, magnetic and thermal properties of $Cu(Rh_{1-x}Pt_x)_2Se_4$ ($0 \leq x \leq 0.35$) are well investigated. $Cu(Rh_{1-x}Pt_x)_2Se_4$ samples in the doping range of $0 \leq x \leq 0.12$ exhibit SC. The optimal doping composition $Cu(Rh_{0.94}Pt_{0.06})_2Se_4$ reaches the maximum of $T_c$ = 3.84 K. In addition, we also investigate the system's magnetic properties and find that the doping of Pt leads to a slight decline to the lower critical magnetic field of the spinel structure material $CuRh_2Se_4$. However, the Pt-doping significantly increases the upper critical magnetic field ($\mu_0H_{c2}$) from 0.6 T of the undoped sample $CuRh_2Se_4$ to 4.93 T of $Cu(Rh_{0.94}Pt_{0.06})Se_4$.

## II. Experimental procedure

Polycrystalline $Cu(Rh_{1-x}Pt_x)_2Se_4$ samples were synthesized by a conventional solid-state reaction method. Stoichiometric mixtures of Cu (99.9%), Rh (99.95%), Pt (> 99.9%), Se ($\geq$ 99.999%) were placed into the quartz tubes. The quartz tubes were then sealed under high vacuum ambiance (< 1 × 10$^{-1}$ MPa) and were heated for ten days in the range of 825 - 860 °C. Subsequently, the obtained powder samples were well reground, pressed into pellets, and sintered again for 48 h at 825 - 860 °C.

Powder X-Ray Diffraction (PXRD), equipped with a Bruker D8 Advance ECO with Cu Kα radiation and an LYNXEYE-XE detector, was employed to define the crystal structure. The acquired XRD data was refined by the software Fullprof Suite in the Rietveld refinement model using Thompson-Cox-Hastings pseudo-Voigt peak shapes. The element ratios were determined by scanning electron microscope combined with energy-dispersive x-ray spectroscopy (SEM-EDS, COXEM EM-30AX). Temperature-dependent resistivity (R-T), temperature-dependent magnetic susceptibility (M-T), and magnetization versus applied magnetic field intensity (M-H) curves were measured by a physical property measurement system (PPMS-14T), which is produced by Quantum Design.

First-principles calculations were performed using density functional theory (DFT) as implemented in Vienna Abinit Simulation Package (VASP) [33]. In particular, the projected augmented wave (PAW) method [34] and Perdew, Burke, and Enzerhoff flavor of generalized gradient approximations [35] to the exchange-correlation functional was employed. The spin-orbit coupling (SOC) interaction was considered in all reported results using a second variational method. The plane-wave basis energy cutoff was chosen to be 540 eV and the Brillouin zone was sampled with 12×12×12 Γ-centered K-mesh to ensure convergence. In order to obtain doping-dependent electronic structure, we construct tight-binding Hamiltonians for both pristine $CuRh_2Se_4$ and pristine $CuPt_2Se_4$ using the maximally projected Wannier function method [36] and symmetrized using full crystal symmetry [37]. The Hamiltonians for doped compounds were then interpolated in the spirit of virtual crystal approximation (VCA).

### III. Results and Discussion

**Figure 1** shows the PXRD patterns of $Cu(Rh_{1-x}Pt_x)_2Se_4$ ($0 \leq x \leq 0.35$) and the relative analyses. The XRD Rietveld refinement for the representative sample $Cu(Rh_{0.95}Pt_{0.05})Se_4$ at room temperature is shown in **Fig. 1a**, which can be well indexed by spinel $CuRh_2Se_4$ (PDF card number: 04-005-8682), while some samples contain a small amount of $RhSe_2$ impurity (less than 5 %, see **Table. S2**). The refined data of other compositions in the system is presented in **Fig. S1** and **Table. S1**. Further, EDXS is employed to confirm the atomic ratio (see **Fig. S2**), implying the obtained compositions are very close to the target compositions. And the EDXS mappings are shown in **Fig. S3**, suggesting homogeneity in polycrystalline samples. When the doping amount of Pt is beyond $x = 0.35$, a large amount of $PtSe_2$ phases appears. As shown in the right side of **Fig. 1b**, the peaks at around 35° are enlarged. A clear left-shift for peaks can be perceived when the doping amount of Pt gradually increases. Based on Braggs Law $2d\sin\theta = n\lambda$, it is not difficult to draw out the explanation that the interplanar distance increases with the doping amount of Pt increasing gradually since the ionic radius of Pt is greater than that of Rh. Subsequently, we performed quantitative analysis on the XRD data by fitting with Fullprof suite using Thompson-Cox-Hastings pseudo-Voigt peak shapes to obtain the lattice parameters. The lattice parameters increases from 10.2645(4)Å for $CuRh_2Se_4$ to 10.3297(4)Å for $Cu(Rh_{0.65}Pt_{0.35})Se_4$. **Figure 1c** shows the lattice parameters as a function of the Pt doping content with a good linearity.

The temperature-dependent resistivity and magnetic susceptibility measurements are carried out to survey the SC. **Figure 2a** displays the normalized resistivity ($\rho/\rho_{300K}$) vs. temperature (**R-T**). For $0 \leq x \leq 0.35$, resistivity decreases as the temperature going down from 300 K to the onset of the zero-drop (around 4 K), implying they are metallic. **Figure 2b** shows the resistivity trend at low temperatures, ranging from 1.8 K to 4.2 K. The superconducting transition width is defined as the temperature difference between 90 % and 10 % resistivity. Resistivity for $Cu(Rh_{1-x}Pt_x)_2Se_4$ ($0 \leq x \leq 0.12$) samples reveals a dramatic decrement that the width is less than 0.3 K. Besides, $T_c$ of each sample, which is defined from the average value of the temperature when the superconducting transition started and ended, is used for establishing the electronic phase diagram. The highest $T_c \approx 3.84$ K in the system is observed at $Cu(Rh_{0.94}Pt_{0.06})_2Se_4$, and is followed by a rapid decline as $x$ increases. Once the doping percentage reaches 15 %, no zero resistivity can be observed down to 1.8 K. In the light of 5 % $RhSe_2$ impurity observed in part of our studied $Cu(Rh_{1-x}Pt_x)_2Se_4$ ($0 \leq x \leq 0.35$) samples, we should examine whether the $T_c$ may originate from the $RhSe_2$ compounds or not. Based on the previous reports, we find out there is no accurate $T_c$ for the undoped $RhSe_2$ compound. Besides, the $T_c$ of $RhSe_{2-x}$ varies irregularly from 1 K to 6 K depending on the Se content [38]. The $T_c$ of the most similar compound $Rh_{0.94}Se_2$ was around 5 K [39]. These facets can ruled out the $T_c$s in $Cu(Rh_{1-x}Pt_x)_2Se_4$ (0 ≤ x ≤ 0.35) are from $RhSe_2$ impurity, combining with the following magnetic and specific heat measurements. The $T_c$ change in $Cu(Rh_{1-x}Pt_x)_2Se_4$ (0 ≤ x ≤ 0.12) could be considered related to the shift of the Fermi level and scattering effect based on the following first-principles calculations and RRR values. When 0 ≤ x ≤ 0.06, $T_c$ increases with the increasing x due to the enhancement of DOS; while x > 0.06, $T_c$ decreases with the increment of x, which may be the disorder effect become more remarkably. Moreover, the residual resistivity ratio (RRR) for each sample is exhibited in **Fig. 2c**. The parent sample exhibits a high RRR value of 37 and a sharp transition to the superconducting state at 3.45 K, denoting that the undoped sample is highly homogeneous [40]. And the subsequent reduction of RRR points out a reinforced effect in electron scattering, which might be related to the rising trend of the upper critical field. Furthermore, SC of $Cu(Rh_{1-x}Pt_x)_2Se_4$ (0 ≤ x ≤ 0.12) is also investigated by magnetic measurements. Zero-field-cooling (ZFC) measurements under 30 Oe magnetic field were applied to detect the diamagnetism of superconducting $Cu(Rh_{1-x}Pt_x)_2Se_4$ (0 ≤ x ≤ 0.12) samples. As shown in **Fig. 2d**, strong diamagnetic shields and clear superconducting transitions are observed. The tendency of $T_c$ shows great agreement with the R-T result at slightly lower level due to the suppression from the applied magnetic field. However, the superconducting transition does not appear as steep as it is expected to be. We suggest that this phenomenon is originated from the decay of the Meissner screen current, which happens commonly in polycrystalline samples [32]. Besides, we employed cylindrical powder samples with an applied magnetic field parallel to the axis of the cylinder for ZFC measurements [41]. Therefore, the demagnetization effect caused the 4πχ values to be less than -1. We then used the formula $\boldsymbol{H_i} = \frac{H_a}{1-n}$ to modify the value of 4πχ, where $\boldsymbol{H_i}$ is the strength of the magnetic field inside the sample and $\boldsymbol{H_a}$ is the applied magnetic field. The values of the demagnetization factor n calculated from geometric elements vary from 0.5924 to 0.6268.

Subsequently, to determine the lower critical transition magnetic field $\mu_0H_{c1}(0)$, we measured the M-H curves at different temperatures. For comparison, the parent sample CuRh$_2$Se$_4$ and the representative sample Cu(Rh$_{0.94}$Pt$_{0.06}$)$_2$Se$_4$ are chosen for the test. The test details of $\mu_0H_{c1}(0)$ are shown in **Fig. 3**. The upper right insets of **Fig. 3a,b** display the M-H curves at various temperatures. When the external magnetic field is weak, the magnetization intensity *M* performs as a linear relation of the external magnetic field H: $M_{fit} = e + fH$, where *e* is the intercept and *f* is the slope of the line [42]. The lower left insets of **Fig. 3a,b** exhibit the $M-M_{fit}$ vs. H curves. Commonly, the value of $\mu_0H_{c1}^*$ is extracted when the difference between *M* and $M_{fit}$ over 1% $M_{max}$. The obtained points shown in the main panel of **Fig. 3a,b** are well complied with the formula $\mu_0H_{c1}(T) = \mu_0H_{c1}(0)(1-(T/T_c)^2)$. The estimated value $\mu_0H_{a1}(0)$ of CuRh$_2$Se$_4$ and Cu(Rh$_{0.94}$Pt$_{0.06}$)$_2$Se$_4$ are identified as 100 Oe, 84 Oe, respectively. Due to demagnetization, the estimated value $\mu_0H_{a1}(0)$ should be modified using the formula $\mu_0H_{c1}(0) = \frac{\mu_0H_{a1}(0)}{1-n}$, where the demagnetization factor *n* of CuRh$_2$Se$_4$ and Cu(Rh$_{0.94}$Pt$_{0.06}$)$_2$Se$_4$ is 0.5461 and 0.5, respectively. The modified lower critical field $\mu_0H_{c1}(0)$ of CuRh$_2$Se$_4$ and Cu(Rh$_{0.94}$Pt$_{0.06}$)$_2$Se$_4$ is 220 Oe and 168 Oe, respectively. This could suggest that Pt substitution can slightly affect the $\mu_0H_{c1}(0)$.

The upper critical fields are studied by the temperature-dependent resistivity measurement under diverse applied fields systematically. **Fig. 4a-f** show the measurement process of Cu(Rh$_{1-x}$Pt$_x$)$_2$Se$_4$ (*x* = 0, 0.05, 0.06). It can be seen in **Fig. 4b, d, f** that $T_c$ puts up a continuous declination with the increase of the applied magnetic field intensity. Data plots of the upper critical field $\mu_0H_{c2}^*$, which are determined by the 50 % criterion of normal-state resistivity values, are used to estimate $\mu_0H_{c2}(0)$ following the Werthamer–Helfand–Hohenburg (WHH) and Ginzberg–Landau (G-L) theories, respectively. $dH_{c2}/dT_c$, which refers to the slope of plots near $T_c$, is used in simplified WHH formula: $\mu_0H_{c2}(0) = -0.693T_c \left(\frac{dH_{c2}}{dT}\right)_{T_c}$. As a dirty limit from the WHH model, the calculated values of CuRh$_2$Se$_4$, Cu(Rh$_{0.95}$Pt$_{0.05}$)$_2$Se$_4$, and Cu(Rh$_{0.94}$Pt$_{0.06}$)$_2$Se$_4$ are 0.81(2) T, 3.22(2) T, 3.75(8) T, respectively. It is worth mentioning that the calculated upper critical field from WHH model must be less than the Pauling limiting field $H^P =$

$1.86T_c$, which is a precondition derived from the Pauling-limiting effect [43]. In this case, $H^P$'s of $CuRh_2Se_4$, $Cu(Rh_{0.95}Pt_{0.05})_2Se_4$, and $Cu(Rh_{0.94}Pt_{0.06})_2Se_4$ are 6.311(2) T, 7.072(2) T, and 7.176(4) T, all of which are higher than the calculated $\mu_0H_{c2}(0)$, respectively. $\mu_0H_{c2}^*$ also follows the function distribution based on G-L theory: $\mu_0H_{c2}(T) = \mu_0H_{c2}(0) * \frac{1-(T/T_c)^2}{1+(T/T_c)^2}$, in which the $\mu_0H_{c2}(0)$ can be calculated. As shown in **Fig. 4a, c, e**, the distribution of spots obeys the function nicely. The estimated $\mu_0H_{c2}(0)$ for $CuRh_2Se_4$, $Cu(Rh_{0.95}Pt_{0.05})_2Se_4$ and $Cu(Rh_{0.94}Pt_{0.06})_2Se_4$ is 1.00(1) T, 4.03(1) T, 4.93(1) T, respectively. Although there is a non-negligible difference between the obtained $\mu_0H_{c2}(0)$ values from the WHH model and G-L theory, the results indicate that Pt substitution can greatly increase the upper critical fields. We propose that the increase of the upper critical field is derived from the effect of Pt substitution. RRR shown in **Fig. 2c** is an indicator of disorder, the reduction of whose value implies the increase in disorder. Once Pt gets in, a sharp drop of RRR value is observed, suggesting that the weak-magnetic element platinum behaves as an effective scattering center. The employment of weak-magnetic element platinum does not affect the critical temperature but significantly augment the disorder. Therefore, the electron scattering is more intense, and the mean free path of carriers declines [44-47]. In addition, from Fig. S4, it can be seen that the enhanced $\mu_0H_{c2}(0)$ is corresponding to the shorter coherence length ($\xi_{GL}$) and reduced residual resistivity ratio (RRR). From these facets, we speculate the enhanced $\mu_0H_{c2}(0)$ in the Pt doped samples may be induced by the shorter coherence length due to the impurity scattering.

Further, the intrinsic property heat capacity measurement at low temperature is performed to confirm the polycrystalline sample $Cu(Rh_{0.92}Pt_{0.09})_2Se_4$ whether it is a bulk superconductor. The analysis of heat capacity measurement is depicted in **Fig. 5**. The heat capacity mainly comes from the contribution of electron ($C_{el.}$) and phonon ($C_{ph.}$), which can be described as $\gamma T$ and $\beta T^3$, respectively. The heat capacity data above $T_c$ can be fitted as the equation $C_p = \gamma T + \beta T^3$. The value of $\gamma$ and $\beta$ are determined to be 22.57(20) mJ·mol$^{-1}$·K$^{-2}$ and 1.65(1) mJ·mol$^{-1}$·K$^{-4}$, respectively. **Fig. 5a** shows the $C_p/T$ vs. $T^2$ curves measured under the magnetic field of 0 T, 1 T and 5 T. It is apparent that the superconducting transition peak shifts to lower temperatures as the magnetic field increases and even disappears under 5 T magnetic field, suggesting CuRh$_{1.82}$Pt$_{0.18}$Se$_4$ is a type-II superconductor. **Fig. 5b** shows the $C_{el.}/T$ vs. $T$ curve in the temperature range between 1.8 K and 5.5 K under zero magnetic fields, where $C_{el.}$ is obtained by the equation $C_{el.} = C_p - \beta T^3$. The estimated $T_c = 3.26$ K determined by an equal-area entropy construction is consistent with the $T_c$'s extracted from resistivity and magnetic susceptibility measurements for the polycrystalline sample Cu(Rh$_{0.92}$Pt$_{0.09}$)$_2$Se$_4$. The normalized specific heat jump value $\Delta C/\gamma T_c$ is calculated to be 1.52, which is slightly higher than the Bardeen-Cooper-Schrieffer (BCS) weak-coupling limit value of 1.43, evidencing bulk SC. The Debye temperature is available in line with the formula $\Theta_D = \left(\frac{12\pi^4 nR}{5\beta}\right)^{1/3}$, where $n$ stands for the account of atom per formula unit and $R$ is the molar gas constant. Under the premise of given Debye temperature and $T_c$, the electron-phonon coupling constant $\lambda_{ep}$ can be calculated to be 0.63 with $\mu^* = 0.13$ by employing the McMillan formula: $\lambda_{ep} = \frac{1.04 + \mu^* \ln\left(\frac{\Theta_D}{1.45 T_c}\right)}{(1-1.62\mu^*)\ln\left(\frac{\Theta_D}{1.45 T_c}\right) - 1.04}$ [48]. And the density of states (DOS) located at the Fermi level [$N(E_F)$] can also be estimated by the formula $N(E_F) = \frac{3}{\pi^2 k_B^2 (1+\lambda_{ep})}\gamma$ with the value of $\gamma$ and $\lambda_{ep}$. The resultant $N(E_F)$ of Cu(Rh$_{0.92}$Pt$_{0.09}$)$_2$Se$_4$ is 5.87 states/eV·f.u., which is lower than that of the undoped sample CuRh$_2$Se$_4$ as shown in **Table 1**. This result also matches the fact that Cu(Rh$_{0.92}$Pt$_{0.09}$)$_2$Se$_4$ has a lower $T_c$ than that of undoped sample CuRh$_2$Se$_4$.

The electronic phase diagram for Cu(Rh$_{1-x}$Pt$_x$)$_2$Se$_4$ ($0 \leq x \leq 0.12$) is performed to summarize our characterizations. The acquired $T_c$ from the experiment result of R-T and M-T is summed up as a curve related to the Pt doping concentration. From **Fig. 6**, we can find that with the increment of Pt substitution, $T_c$ shows an upward trend and reaches the maximum value of 3.85 K for Cu(Rh$_{0.94}$Pt$_{0.06}$)$_2$Se$_4$. Subsequently, $T_c$ drops down at the higher Pt doping region. It can be taken as evidence that Pt chemical doping can be used to tune the SC for the spinel CuRh$_2$Se$_4$.

In order to understand the above experimental results, we have also performed first-principles calculations. **Fig. 7a** shows the electronic band structure of pristine

CuRh$_2$Se$_4$ and CuPt$_2$Se$_4$. For CuRh$_2$Se$_4$ (**Fig. 7a**, left panel), two doubly degenerate bands (band No. 89 – 92) cross the Fermi level once SOC interaction is considered, creating 3 doubly degenerate Fermi surface sheets including a pair of drum-like pockets and a pair of cross-like pockets at the zone boundaries, as well as a pair of star-like pockets around Γ point (**Fig. 7b**). Interestingly, these two doubly degenerate bands are well separated from all other band states, and are dominated by the Se-4$p$ and Rh-4$d$ orbitals, as also evident from the DOS calculation (**Fig. 7c**). In general, the CuPt$_2$Se$_4$ band structure (**Fig. 7a**, right panel) resembles a heavily electron-doped CuRh$_2$Se$_4$, since the top of band 91/92 is now ~0.6 eV below the Fermi level. However, there are several important differences, which invalidate the commonly adopted rigid-band shifting method. Firstly, bands 89-92 are now entangled with the bands below them. Secondly, bands 89-92 became more dispersive than they were in CuRh$_2$Se$_4$, possibly due to the larger hopping terms between Pt-5$d$ orbitals and Se-4$p$ orbitals than those between Rh-4$d$ and Se-4$p$, as a result of the larger atomic radius of 5$d$ wave functions. As a result, although a direct gap is always present between bands 91/92 and 93/94, the indirect gap between them is decreased from 247 meV to -30 meV.

With the electronic structure of both pristine compound explained, we now address the electronic structure of Cu(Rh$_{1-x}$Pt$_x$)$_2$Se$_4$ under the spirit of VCA. The doping-dependent DOS is shown in **Fig. 7d**. The elemental substitution causes a significant red-shift of the DOS peak around 0.1 eV above the Fermi level, corresponding to the electron doping effect. At $x = 0.1$, the DOS peak shifts very close to the Fermi level, resulting in a peak of $N(E_F)$ [inset of **Fig. 7d**]. Increasing Pt doping to $x > 0.1$, the DOS peak moves away from the Fermi level, and $N(E_F)$ decreases monotonically [inset of **Fig. 7d**]. In addition, it is apparent that the van Hove singularity close to the band edge (located approximately 0.2 eV above the Fermi level) for pristine CuRh$_2$Se$_4$ is quickly suppressed upon Pt doping. This can be understood since the Pt-5$d$ orbital radius is much larger than the Rh-4d orbital radius. Therefore, the enhanced hoppings effectively enhance the three-dimensionality. According to the BCS theory $T_c \propto exp\left(-\frac{1}{N(E_F)V}\right)$, where $V$ is the attractive potential, the $N(E_F)$ change with respect to doping content $x$ is roughly consistent with the experimental observation. However, the existence of RhSe$_2$ plays a role of hole dopant. Based on the monovalence assumption of copper, the absent of copper in the impurity produce 0.5 hole/f.u., causing an additional blue-shift depending on the concentration of RhSe$_2$ impurity. As a result, the measured peak was discovered at a lower doping level than the calculated one.

## IV. Conclusion

In summary, we have succeeded in synthesizing a series of Cu(Rh$_{1-x}$Pt$_x$)$_2$Se$_4$ ($0 \leq x \leq 0.35$) spinels via a conventional solid-state reaction. Combining the experimental and theoretical studies, we find that Pt substitution plays a positive role in $T_c$ enhancement, reaching a maximum $T_c$ of 3.85 K in the optimal doping composition Cu(Rh$_{0.94}$Pt$_{0.06}$)$_2$Se$_4$. Meanwhile, it can vastly raise the upper critical magnetic field $\mu_0 H_{c2}(0)$ by the disorder. This finding shed light on the discovery of new spinel SC materials and provide a new material platform to study the mechanism of spinel superconductors.

## Acknowledgments

This work is supported by the National Natural Science Foundation of China (Grants No. 11922415, 11874137), Guangdong Basic and Applied Basic Research Foundation (2019A1515011718), Key Research & Development Program of Guangdong Province, China (2019B110209003), and the Pearl River Scholarship Program of Guangdong Province Universities and Colleges (20191001). The calculations were performed on the High Performance Computing Center at School of Physics, Hangzhou Normal University.

## V. Reference

[1] J. Hemberger, H. A. K. von Nidda, V. Tsurkan, and A. Loidl, Phys. Rev. Lett. **98**, 147203 (2007).

[2] A. P. Ramirez, R. J. Cava, and J. Krajewski, Nature **386**, 156 (1997).

[3] Z. Yang, S. Tan, Z. Chen, and Y. Zhang, Phys. Rev. B **62**, 13872 (2000).

[4] S. Weber, P. Lunkenheimer, R. Fichtl, J. Hemberger, V. Tsurkan, and A. Loidl, Phys. Rev. Lett. **96**, 157202 (2006).

[5] K. Ohgushi, T. Ogasawara, Y. Okimoto, S. Miyasaka, and Y. Tokura, Phys. Rev. B **72**, 155114 (2005).

[6] J. Hemberger, P. Lunkenheimer, R. Fichtl, H. A. Krug von Nidda, V. Tsurkan, and A. Loidl, Nature **434**, 364 (2005).

[7] D. C. Johnson, J. Low Temp. Phys., **25**, 145-175 (1976)

[8] W. Hu, Z. Feng, B. Gong, G. He, D.Li, M. Qin, Y. Shi, Q. Li, Q. Zhang, J. Yuan, B. Zhu, K. Liu, T. Xiang, L. Gu, F. Zhou, X. Dong, Z. Zhao, K. Jin, Phys. Rev. B **101**, 6, 220510(R) (2020).

[9] T. Bitoh, T. Hagino, Y. Seki, S. Chikazawa, and S. Nagata, J. Phys. Soc. Japan **61**, 3011 (1992).

[10] M. Ito, J. Hori. H. Kurisaki, H. Okada, A. J. Perez Kuroki, N. Ogita, M. Udagawa, H. Fujii, F. Nakamura, T. Fujita, and T. Suzuki, Phys. Rev. Lett. **91**, 077001 (2003).

[11] M. Ito, A. Taira, and K. Sonoda, Acta Phys. Pol. A **131**, 1450 (2017).

[12] T. Shirane, T. Hagino, Y. Seki, T. Bitoh, S. Chikazawa, and S. Nagata, J. Phys. Soc. Japan **62**, 374 (1993).

[13] R. N. Shelton, D. C. Johnston, and H. Adrian, Solid State Commun. **20**, 1077 (1976).

[14] T. Hagino, Y. Seki, N. Wada, S. Tsuji, T. Shirane, K. Kumagai, and S. Nagata, Phys. Rev. B **51**, 12673 (1995).

[15] H. Suzuki, T. Furubayashi, G. Cao, H. Kitazawa, A. Kamimura, K. Hirata, and T. Matsumoto, J. Phys. Soc. Japan **68**, 2495 (1999).

[16] G. Cao, H. Kitazawa, H. Suzuki, T. Furubayashi, K. Hirata, and T. Matsumoto, Physica C Supercond. **341-348**, 735 (2000).

[17] H. Luo, T. Klimczuk, L. Müchler, L. Schoop, D. Hirai, M. K. Fuccillo, C. Felser, and R. J. Cava, Phys. Rev. B **87**, 214510 (2013).

[18] Y. Y. Jin, S. H. Sun, Y. W. Cui, Q. Q. Zhu, L. W. Ji, Z. Ren, and G. H. Cao, Phys. Rev. Mater. **5**, 7, 074804 (2021).

[19] T. Furubayashi, T. Matsumoto, T. Hagino, and S. Nagata, J. Phys. Soc. Japan **63**, 3333 (1994).

[20] S. Nagata, Chinese J. Phys. **43**, 722 (2005).

[21] G. Oomi, T. Kagayama, I. Yoshida, T. Hagino, and S. Nagata, J. Magn. Magn. Mater. **140-144**, 157 (1995).

[22] M. Boubeche, J. Yu, C. Li, H. Wang, L. Zeng, Y. He, X. Wang, W. Su, M. Wang, D. Yao, Z. Wang, H. Luo, Chinese Physics Letters **38**, 037401 (2021).

[23] Y. Qi, Z. Gao, L. Wang, D. Wang, X. Zhang, C. Yao, C. Wang, C. Wang, and Y. Ma, EPL (Europhysics Letters) **97**, 17008 (2012).

[24] M. Robbins, R. H. Willens, and R. C. Miller, Solid State Commun. **5**, 933 (1967).

[25] M. Ito, K. Ishii, F. Nakamura, and T. Suzuki, Physica B Condens. Matter **359-361**, 1198 (2005).

[26] M. I. Kholil and M. T. H. Bhuiyan, Results Phys. **12**, 73 (2019).

[27] N. Ni, M. E. Tillman, J. Q. Yan, A. Kracher, S. T. Hannahs, S. L. Bud'ko, and P. C. Canfield, Phys. Rev. B **78**, 214515 (2008).

[28] C. Wang, Y. Li, Z. Zhu, S. Jiang, X. Lin, Y. Luo, S. Chi, L. Li, Z. Ren, M. He, H. Chen, Y. Wang, Q. Tao, G. Cao, Z. Xu, Phys. Rev. B **79**, 054521 (2009).

[29] L. Li, Y. Luo, Q. Wang, H. Chen, Z. Ren, Q. Tao, Y. Li, X. Lin, M. He, Z. Zhu, G. Cao, Z. Xu, New J. Phys. **11**, 025008 (2009).

[30] F. Han, X. Zhu, P. Cheng, G. Mu, Y. Jia, L. Fang, Y. Wang, H. Luo, B. Zeng, B. Shen, L. Shan, C. Ren, H. Wen, Phys. Rev. B **80**, 024506 (2009).

[31] S. Sharma, A. Bharathi, S. Chandra, V. R. Reddy, S. Paulraj, A. T. Satya, V. S. Sastry, A. Gupta, and C. S. Sundar, Phys. Rev. B **81**, 174512 (2010).

[32] X. Zhu, F. Han, G. Mu, P. Cheng, J. Tang, J. Ju, K. Tanigaki, and H. Wen, Phys. Rev. B **81**, 104525 (2010).

[33] G. Kresse, J. Non-Cryst. Solids (Netherlands) **192-193**, 222 (1995).

[34] G. Kresse and D. Joubert, Phys. Rev. B **59**, 1758 (1999).

[35] J. P. Perdew, K. Burke, and M. Ernzerhof, Phys. Rev. Lett. **77**, 3865 (1996).

[36] N. Marzari and D. Vanderbilt, Phys. Rev. B, Condens. Matter (USA) **56**, 12847 (1997).

[37] G. Zhi, C. Xu, S. Wu, F. Ning, C. Cao, Comp. Phys. Commun. 271, 108196 (2022)

[38] B. T. Matthias, E. Corenzwit, and C. E. Miller, Phys. Rev. 93, 1415 (1954).

[39] J. Guo, Y. Qi, S. Matsuishi, H. Hosono, J. Am. Chem. Soc. 134, 20001-20004 (2012)

[40] US Department of Commerce, J. Res. Natl. Inst. Stan. **116**, 489 (2011).

[41] R. Prozorov, V. G. Kogan, Phys. Rev. Appl. 10(1), 014030 (2018).

[42] H. Luo, W. Xie, J. Tao, I. Pletikosic, T. Valla. G. S. Sahasrabudhe, G. Osterhoudt, E. Sutton, K. S. Burch, E. M. Seibel, J. W. Krizan, Y. Zhu, R. J. Cava, Chem. Mat. **28**, 1927 (2016).

[43] A. M. Clogston, Phys. Rev. Lett. **9**, 266 (1962).

[44] X. L. Wang, S. X. Dou, M. Hossain, Z. X. Cheng, and T. Silver, Phys. Rev. B **81** (2010).

[45] K. S. B. De Silva, X. Xu, X. L. Wang, D. Wexler, D. Attard, F. Xiang, and S. X. Dou, Scripta Materialia **67**, 802 (2012).

[46] H. T. Wang, L. J. Li, D. S. Ye, X. H. Cheng, and Z. A. Xu, Chinese Phys. B **16**, 2471 (2007).

[47] M. X. Liu, Z. Z. Gan, Chinese Phys. B **16**, 826 (2007).

[48] W. L. McMillan, Phys. Rev. **167**, 331 (1968).

**Figures caption:**

**Fig. 1** Structural characterizations of Cu(Rh$_{1-x}$Pt$_x$)$_2$Se$_4$ ($0 \leq x \leq 0.35$). (a) The XRD results after Rietveld refinement for Cu(Rh$_{0.95}$Pt$_{0.05}$)$_2$Se$_4$; (b) XRD patterns for Cu(Rh$_{1-x}$Pt$_x$)$_2$Se$_4$ ($0 \leq x \leq 0.35$). The peak representing the lattice plane (400) was enlarged in the inset. (c) The trend of lattice parameter as the change of the Pt doping amount.

**Fig. 2** The measurements of temperature-dependent resistivity and magnetic susceptibility for Cu(Rh$_{1-x}$Pt$_x$)$_2$Se$_4$ ($0 \leq x \leq 0.35$). (a) The overview of the normalized resistivity ($\rho/\rho_{300K}$) in the range of 1.8 K to 300 K; (b) The temperature dependence of normalized resistivity ($\rho/\rho_{300K}$) in the range of 1.8 K to 4.5 K; (c) The RRR value vs. Pt content; (d) The ZFC M-T curves for Cu(Rh$_{1-x}$Pt$_x$)$_2$Se$_4$ ($0 \leq x \leq 0.12$) in the range of 1.8 to 4.2 K.

**Fig. 3** The temperature dependence measurement of lower critical ($\mu_0 H_{c1}$) field for CuRh$_2$Se$_4$ and Cu(Rh$_{0.94}$Pt$_{0.06}$)$_2$Se$_4$. The inset located in the upper-right corner showed the variation of magnetization intensity with the increment of the applied magnetic field. The inset located in the low left quarter showed the intersection point of $M-M_{fit}$ and 1% $M_{max}$. The curves in the insets of Figure are taken by applying the field after zero-field cooling.

**Fig. 4** The temperature dependence measurement of the upper critical ($\mu_0 H_{c2}$) fields for Cu(Rh$_{1-x}$Pt$_x$)$_2$Se$_4$ ($x = 0, 0.05, 0.06$); (a), (c) and (e) show the refinements of Cu(Rh$_{1-x}$Pt$_x$)$_2$Se$_4$ ($x = 0, 0.05, 0.06$), respectively. The red curve exhibits the refinement by G-L theory, while the blue curve displays the refinement by WHH model; (b), (d) and (f) show the detailed process for determining $\mu_0 H_{c2}^*$.

**Fig. 5** The temperature-dependent specific heat for Cu(Rh$_{0.91}$Pt$_{0.09}$)$_2$Se$_4$. (a) Specific heat divided by temperature $C_p/T$ vs. squared temperature $T^2$ under diverse applied magnetic fields. (b) Electronic contribution to the heat capacity divided by temperature

($C_{el.}/T$) vs. temperature $T$ in the range of 2.0 to 3.8 K without an applied magnetic field.

**Fig. 6** The electronic phase diagram for Cu(Rh$_{1-x}$Pt$_x$)$_2$Se$_4$ ($0 \leq x \leq 0.12$) vs Pt content.

**Fig. 7** The electronic structure of Cu(Rh$_{1-x}$Pt$_x$)$_2$Se$_4$. (a) Band structure of CuRh$_2$Se$_4$ ($x$ = 0) (left panel) and CuPt$_2$Se$_4$ ($x$ = 1) (right panel). The width of bands are proportional to the orbital composition (red: Rh-4$d$/Pt-5$d$, blue: Se-4$p$, green: Cu-3$d$). (b) Fermi surface sheets of CuRh$_2$Se$_4$. (c) Total and projected DOS of CuRh$_2$Se$_4$. (d) Total DOS of doped system.

**Table. 1** The comparison of superconducting characteristic parameters for spinel compound.

**Table. S1** Structural parameters derived from Rietveld refinement of Cu(Rh$_{1-x}$Pt$_x$)$_2$Se$_4$. Space group Fd$\bar{3}$m (No. 227).

**Table. S2** The proportion of each phase derived from refinement using RIR method (RIR$_{CuRh2Se4}$ = 6.96, RIR$_{RhSe2}$ = 5.32).

**Fig. S1** The XRD result after Rietveld refinement for (a) CuRh$_2$Se$_4$, (b) Cu(Rh$_{0.985}$Pt$_{0.015}$)$_2$Se$_4$, (c) Cu(Rh$_{0.98}$Pt$_{0.02}$)$_2$Se$_4$, (d) Cu(Rh$_{0.96}$Pt$_{0.04}$)$_2$Se$_4$, (e) Cu(Rh$_{0.94}$Pt$_{0.06}$)$_2$Se$_4$, (f) Cu(Rh$_{0.93}$Pt$_{0.07}$)$_2$Se$_4$, (g) Cu(Rh$_{0.91}$Pt$_{0.09}$)$_2$Se$_4$, (h) Cu(Rh$_{0.88}$Pt$_{0.12}$)$_2$Se$_4$.

**Fig. S2** The EDXS spectrum of (a) CuRh$_2$Se$_4$, (b) Cu(Rh$_{0.95}$Pt$_{0.05}$)$_2$Se$_4$, (c) Cu(Rh$_{0.94}$Pt$_{0.06}$)$_2$Se$_4$, (d) Cu(Rh$_{0.93}$Pt$_{0.07}$)$_2$Se$_4$ and the obtained percentage of atoms.

**Fig. S3** EDXS mappings of Cu(Rh$_{1-x}$Pt$_x$)$_2$Se$_4$ ($x$ = 0, 0.05, 0.06, 0.07).

**Fig. S4** The relationship between upper critical fields ($\mu_0H_{c2}$) and coherence length ($\xi_{GL}$) and reduced residual resistivity ratio (RRR).

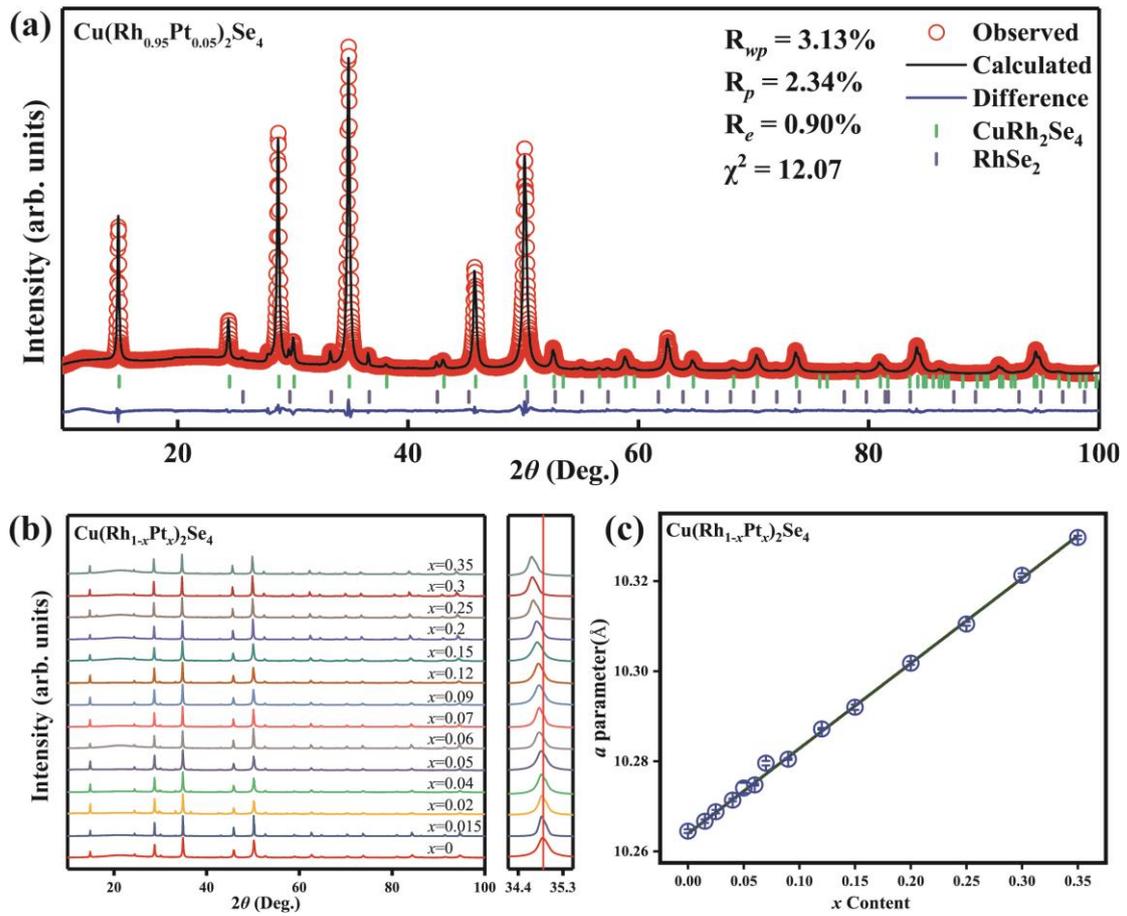

**Fig. 1** Structural characterizations of Cu(Rh$_{1-x}$Pt$_x$)$_2$Se$_4$ ($0 \leq x \leq 0.35$). (a) The XRD result after Rietveld refinement for Cu(Rh$_{0.95}$Pt$_{0.05}$)$_2$Se$_4$; (b) XRD patterns for Cu(Rh$_{1-x}$Pt$_x$)$_2$Se$_4$ ($0 \leq x \leq 0.35$). The peak representing the lattice plane (400) was enlarged in the inset. (c) The trend of lattice parameter as the change of the Pt doping amount.

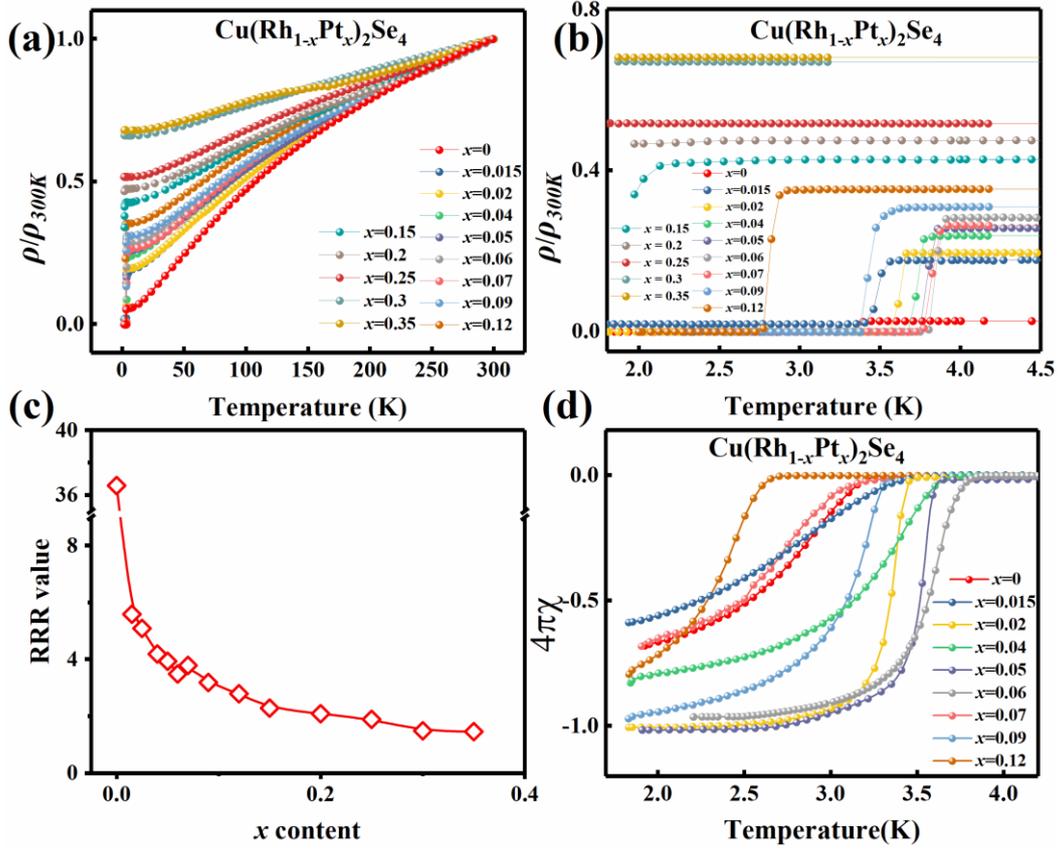

**Fig. 2** The measurements of temperature-dependent resistivity and magnetic susceptibility for Cu(Rh$_{1-x}$Pt$_x$)$_2$Se$_4$ ($0 \leq x \leq 0.35$). (a) The overview of the normalized resistivity ($\rho/\rho_{300K}$) in the range of 1.8 K to 300 K; (b) The temperature dependence of normalized resistivity ($\rho/\rho_{300K}$) in the range of 1.8 K to 4.5 K; (c) The RRR value vs. Pt content; (d) The ZFC M-T curve for Cu(Rh$_{1-x}$Pt$_x$)$_2$Se$_4$ ($0 \leq x \leq 0.35$) in the range of 1.8 to 4.2 K.

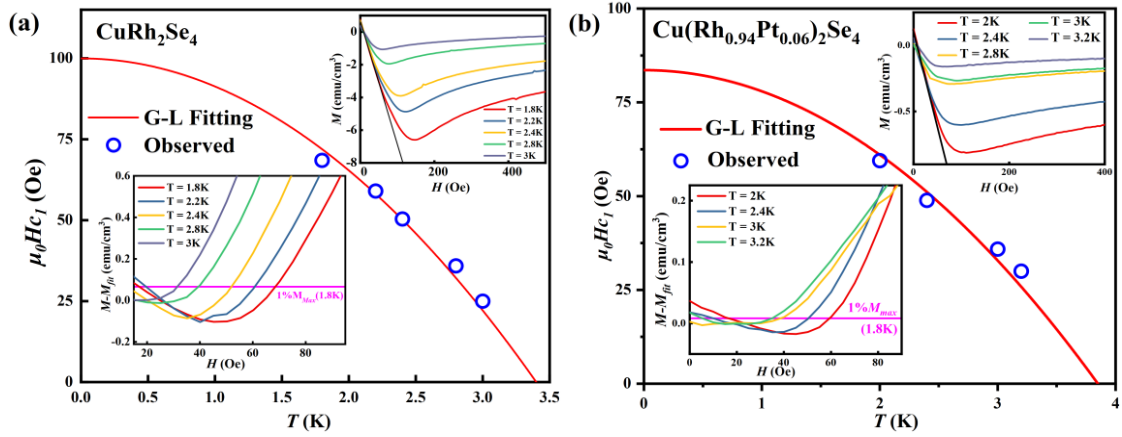

**Fig. 3** The temperature dependence measurement of lower critical ($\mu_0H_{c1}$) field for $CuRh_2Se_4$ and $Cu(Rh_{0.94}Pt_{0.06})_2Se_4$. The inset located in the upper-right corner showed the variation of magnetization intensity with the increment of the applied magnetic field. The inset located in the low left quarter showed the intersection point of $M-M_{fit}$ and 1% $M_{max}$. The curves in the insets of Figure are taken by applying the field after zero-field cooling.

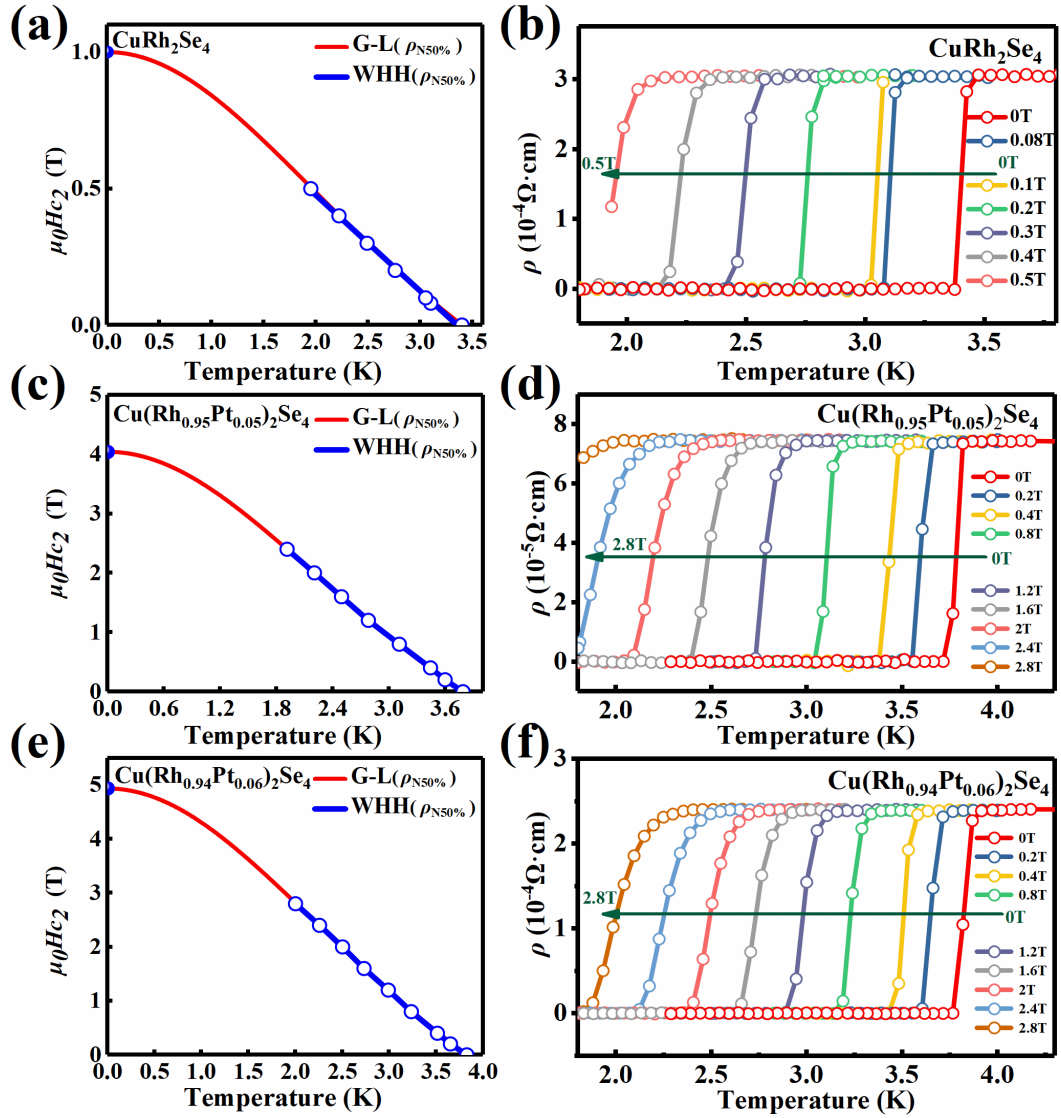

**Fig. 4** The temperature dependence measurement of the upper critical ($\mu_0H_{c2}$) fields for Cu(Rh$_{1-x}$Pt$_x$)$_2$Se$_4$ ($x$ = 0, 0.05, 0.06); (a), (c) and (e) show the refinements of Cu(Rh$_{1-x}$Pt$_x$)$_2$Se$_4$ ($x$ = 0, 0.05, 0.06), respectively. The red curve exhibits the refinement by G-L theory, while the blue curve displays the refinement by WHH model; (b), (d) and (f) show the detailed process for determining $\mu_0H_{c2}^*$.

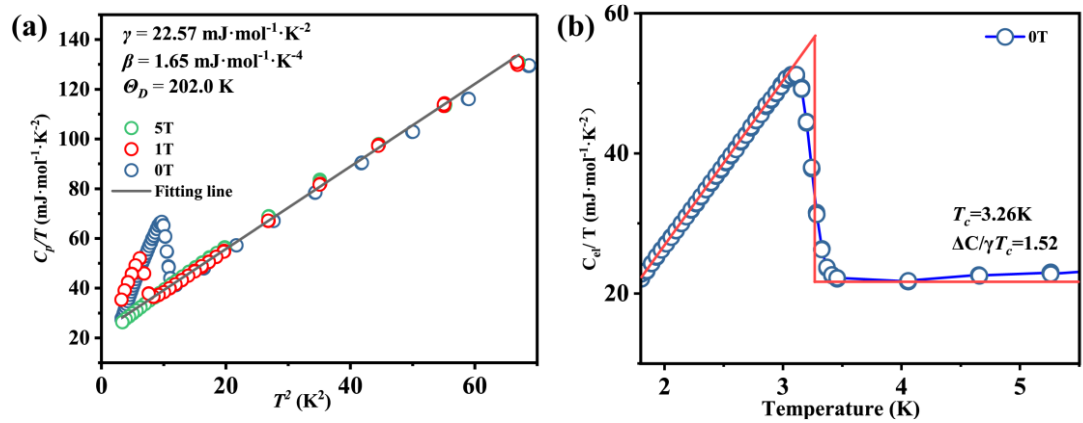

**Fig. 5** The temperature-dependent specific heat for Cu(Rh$_{0.91}$Pt$_{0.09}$)$_2$Se$_4$. (a) Specific heat divided by temperature $C_p/T$ vs. squared temperature $T^2$ under diverse applied magnetic fields. (b) Electronic contribution to the heat capacity divided by temperature ($C_{el.}/T$) vs. temperature $T$ in the range of 2.0 to 3.8 K without an applied magnetic field.

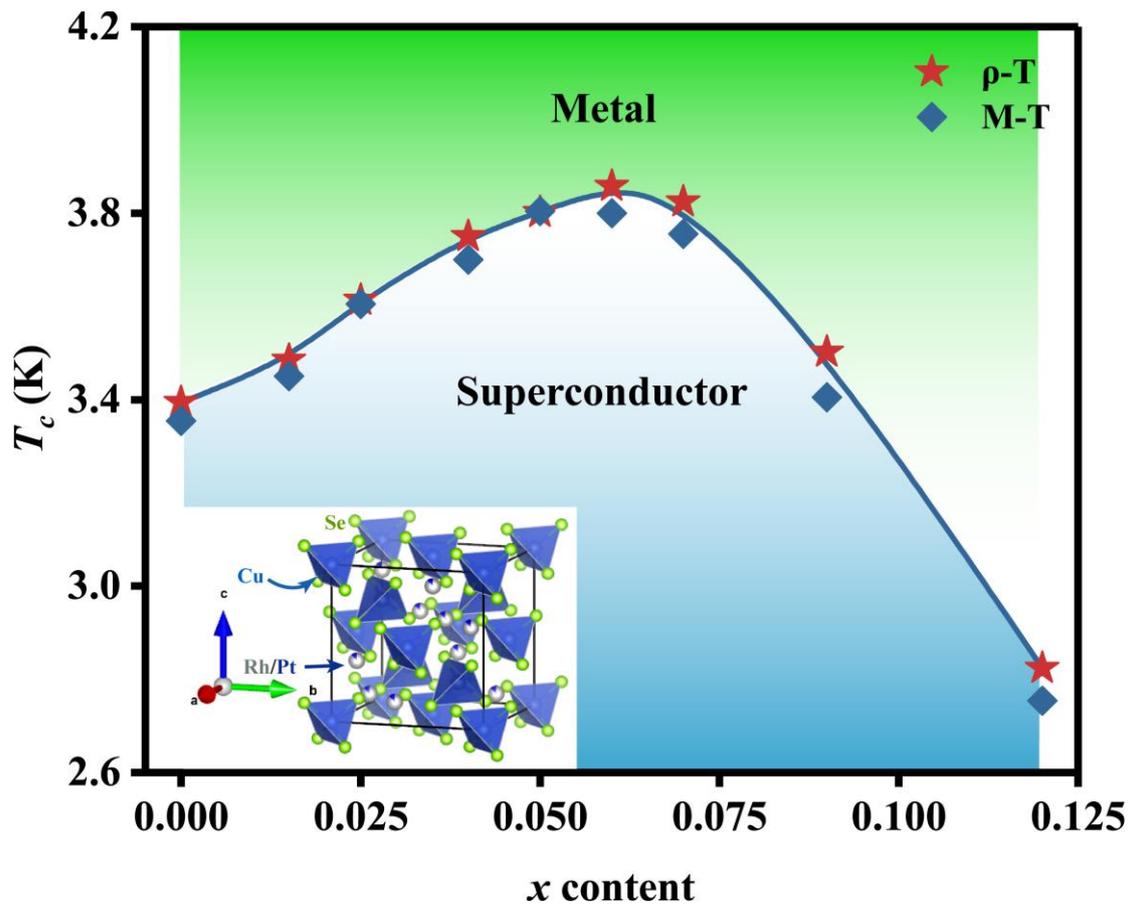

**Fig. 6** The electronic phase diagram for Cu(Rh$_{1-x}$Pt$_x$)$_2$Se$_4$ ($0 \leq x \leq 0.12$) vs Pt content.

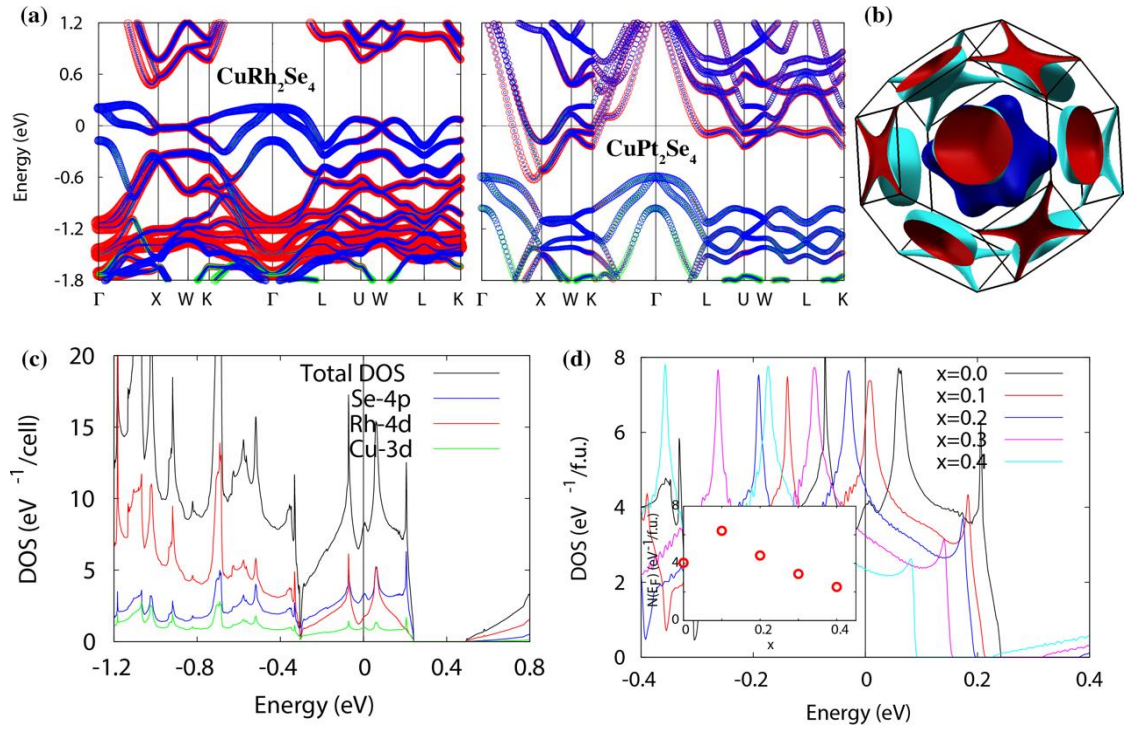

**Fig. 7** The electronic structure of $Cu(Rh_{1-x}Pt_x)_2Se_4$. (a) Band structure of $CuRh_2Se_4$ ($x = 0$) (left panel) and $CuPt_2Se_4$ ($x = 1$) (right panel). The width of bands are proportional to the orbital composition (red: Rh-4$d$/Pt-5$d$, blue: Se-4$p$, green: Cu-3$d$). (b) Fermi surface sheets of $CuRh_2Se_4$. (c) Total and projected DOS of $CuRh_2Se_4$. (d) Total DOS of doped system.

**Table. 1** The comparison of superconducting characteristic parameters for spinel compounds. $\mu_0H^P$ is the Pauling limited field. $\xi_{GL}$ is the Ginzburg-Laudau coherence length at 0 K. $\gamma$ is the constant of electronic specific heat and $\beta$ is the phonon contribution terms. $\Theta_D$ is the Debye temperature. $\Delta C/\gamma T_c$ is the normalized specific heat jump value. $\lambda_{ep}$ is the electron-phonon coupling constant. $N(E_F)$ is the DOS located at the Fermi level.

| Compound | CuRh$_2$Se$_4$ [12] | Cu(Rh$_{0.95}$Pt$_{0.05}$)$_2$Se$_4$ | Cu(Rh$_{0.94}$Pt$_{0.06}$)$_2$Se$_4$ | Cu(Rh$_{0.91}$Pt$_{0.09}$)$_2$Se$_4$ | Cu(Ir$_{0.8}$Pt$_{0.2}$)$_2$Se$_4$ [17] | CuRh$_2$S$_4$ [12] |
|---|---|---|---|---|---|---|
| $T_c$ (K) | 3.38(1) | 3.80(1) | 3.84(2) | 3.47(3) | 1.76 | 3.5 |
| $\mu_0H_{c1}(0)$ (Oe) | 220(6) | - | 168(12) | - | - | - |
| $\mu_0H_{c2}(0)$ (T) ($\rho^{50\%}{}_N$ G-L theory) | 1.00(1) | 4.03(1) | 4.93(1) | - | - | - |
| -d$H_{c2}$/d$T_c$ (T/K) | 0.352(8) | 1.233(8) | 1.418(31) | - | 2.6 | 0.614 |
| $\mu_0H_{c2}(0)$ (T) ($\rho^{50\%}{}_N$ WHH theory) | 0.81(2) | 3.22(2) | 3.75(8) | - | 3.2 | 2.0 |
| $\mu_0H^P$ (T) | 6.311(2) | 7.072(2) | 7.176(4) | - | 3.2 | 8.74 |
| $\xi_{GL}$ (Å) ($\rho^{50\%}{}_N$ WHH theory) | 20.1 | 10.1 | 9.36 | - | 101 | - |
| $\gamma$ (mJ·mol$^{-1}$·K$^{-2}$) | 21.4 | - | - | 22.57(20) | 16.5 | 26.9 |
| $\beta$ (mJ·mol$^{-1}$·K$^{-4}$) | - | - | - | 1.65(1) | 1.41 | - |
| $\Theta_D$ (K) | 218 | - | - | 202.0 | 212 | 258 |
| $\Delta C/\gamma T_c$ | 1.68 | - | - | 1.52 | 1.58 | 1.89 |
| $\lambda_{ep}$ | 0.63 | - | - | 0.63 | 0.57 | 0.66 |
| $N(E_F)$ (states/eV f.u.) | - | - | - | 5.87 | 4.45 | - |
| $\zeta$ (nm) | 18.2 | 8.89 | 8.16 | - | 10.1 | |

# Supporting Information

# Superconductivity with the enhanced upper critical field in the Pt-Doping CuRh$_2$Se$_4$ spinel

*Yiyi He[1,#], Yi-Xin You[2,#], Lingyong Zeng[1], Shu Guo[3,4], Huawei Zhou[1], Kuan Li[1], Yanhao Huang[1], Peifeng Yu[1], Chao Zhang[1], Chao Cao[2,5], Huixia Luo[1]\**

[1] School of Materials Science and Engineering, State Key Laboratory of Optoelectronic Materials and Technologies, Key Lab of Polymer Composite & Functional Materials, Guangzhou Key Laboratory of Flexible Electronic Materials and Wearable Devices, Sun Yat-Sen University, No. 135, Xingang Xi Road, Guangzhou, 510275, P. R. China.

[2] School of Physics and Hangzhou Key Laboratory of Quantum Matters, Hangzhou Normal University, Hangzhou, 311121, P. R. China Shenzhen Institute for Quantum

[3] Shenzhen Institute for Quantum Science and Engineering (SIQSE), Southern University of Science and Technology, Shenzhen, P. R. China

[4] International Quantum Academy (SIQA), and Shenzhen Branch, Hefei National Laboratory, Futian District, Shenzhen, P. R. China

[5] Department of Physics and Center for Correlated Matter, Zhejiang University, Hangzhou, 310013, P. R. China

[#] These authors contributed equally to this work.

\* *Corresponding author/authors complete details (Telephone; E-mail:) (+86)-2039386124; E-mail address: luohx7@mail.sysu.edu.cn*

Table. S1 Structural parameters derived from Rietveld refinement of Cu(Rh$_{1-x}$Pt$_x$)$_2$Se$_4$. Space group Fd$\bar{3}$m (No. 227).

| **CuRh$_2$Se$_4$** | \multicolumn{6}{l}{$R_{wp}$ = 5.66%, $R_p$ = 4.02%, $R_{exp}$ = 1.63%, $\chi^2$ = 12.05} |
|---|---|---|---|---|---|---|
| Label | x | y | z | site | Occupancy | Multiplicity |
| Cu | 0 | 0 | 0 | 8a | 1.000(7) | 8 |
| Rh | 0.625 | 0.625 | 0.625 | 16d | 0.945(5) | 16 |
| Pt | 0.625 | 0.625 | 0.625 | 16d | 0 | 16 |
| Se | 0.383730(8) | 0.383730(8) | 0.383730(8) | 32e | 1.000(4) | 32 |
| **Cu(Rh$_{0.985}$Pt$_{0.015}$)$_2$Se$_4$** | \multicolumn{6}{l}{$R_{wp}$ = 5.87%, $R_p$ = 3.85%, $R_{exp}$ = 1.19%, $\chi^2$ = 24.33} |
| Label | x | y | z | site | Occupancy | Multiplicity |
| Cu | 0 | 0 | 0 | 8a | 1.000(5) | 8 |
| Rh | 0.625 | 0.625 | 0.625 | 16d | 0.959(4) | 16 |
| Pt | 0.625 | 0.625 | 0.625 | 16d | 0.00(13) | 16 |
| Se | 0.383462(7) | 0.383462(7) | 0.383462(7) | 32e | 1.000(4) | 32 |
| **Cu(Rh$_{0.98}$Pt$_{0.02}$)$_2$Se$_4$** | \multicolumn{6}{l}{$R_{wp}$ = 3.19%, $R_p$ = 2.41%, $R_{exp}$ = 1.31%, $\chi^2$ = 5.90} |
| Label | x | y | z | site | Occupancy | Multiplicity |
| Cu | 0 | 0 | 0 | 8a | 1.000(4) | 8 |
| Rh | 0.625 | 0.625 | 0.625 | 16d | 0.961(4) | 16 |
| Pt | 0.625 | 0.625 | 0.625 | 16d | 0.00(3) | 16 |
| Se | 0.38315(5) | 0.38315(5) | 0.38315(5) | 32e | 0.978(8) | 32 |
| **Cu(Rh$_{0.96}$Pt$_{0.04}$)$_2$Se$_4$** | \multicolumn{6}{l}{$R_{wp}$ = 4.38%, $R_p$ = 3.32%, $R_{exp}$ = 1.14%, $\chi^2$ = 14.67} |
| Label | x | y | z | site | Occupancy | Multiplicity |
| Cu | 0 | 0 | 0 | 8a | 1.000(3) | 8 |
| Rh | 0.625 | 0.625 | 0.625 | 16d | 0.876(2) | 16 |
| Pt | 0.625 | 0.625 | 0.625 | 16d | 0.040(7) | 16 |
| Se | 0.38252(5) | 0.38252(5) | 0.38252(5) | 32e | 1.00(2) | 32 |
| **Cu(Rh$_{0.95}$Pt$_{0.05}$)$_2$Se$_4$** | \multicolumn{6}{l}{$R_{wp}$ = 3.33%, $R_p$ = 2.51%, $R_{exp}$ = 0.9%, $\chi^2$ = 13.59} |
| Label | x | y | z | site | Occupancy | Multiplicity |
| Cu | 0 | 0 | 0 | 8a | 0.998(4) | 8 |

| Label | x | y | z | site | Occupancy | Multiplicity |
|---|---|---|---|---|---|---|
| Rh | 0.625 | 0.625 | 0.625 | 16d | 0.830(2) | 16 |
| Pt | 0.625 | 0.625 | 0.625 | 16d | 0.0499(15) | 16 |
| Se | 0.382909(3) | 0.382909(3) | 0.382909(3) | 32e | 0.95(3) | 32 |

| **Cu(Rh$_{0.94}$Pt$_{0.06}$)$_2$Se$_4$** | $R_{wp}$ = 6.67%, $R_p$ = 4.64%, $R_{exp}$ = 1.67%, $\chi^2$ = 15.89 | | | | | |
|---|---|---|---|---|---|---|
| Label | x | y | z | site | Occupancy | Multiplicity |
| Cu | 0 | 0 | 0 | 8a | 1.000(4) | 8 |
| Rh | 0.625 | 0.625 | 0.625 | 16d | 0.866(3) | 16 |
| Pt | 0.625 | 0.625 | 0.625 | 16d | 0.057(15) | 16 |
| Se | 0.381441(3) | 0.381441(3) | 0.381441(3) | 32e | 1.000(19) | 32 |

| **Cu(Rh$_{0.93}$Pt$_{0.07}$)$_2$Se$_4$** | $R_{wp}$ = 8.30%, $R_p$ = 6.23%, $R_{exp}$ = 1.54%, $\chi^2$ = 29.14 | | | | | |
|---|---|---|---|---|---|---|
| Label | x | y | z | site | Occupancy | Multiplicity |
| Cu | 0 | 0 | 0 | 8a | 1.000(4) | 8 |
| Rh | 0.625 | 0.625 | 0.625 | 16d | 0.838(3) | 16 |
| Pt | 0.625 | 0.625 | 0.625 | 16d | 0.0710(18) | 16 |
| Se | 0.38332(3) | 0.38332(3) | 0.38332(3) | 32e | 1.000(3) | 32 |

| **Cu(Rh$_{0.91}$Pt$_{0.09}$)$_2$Se$_4$** | $R_{wp}$ = 4.97%, $R_p$ = 3.50%, $R_{exp}$ = 1.11%, $\chi^2$ = 20.04 | | | | | |
|---|---|---|---|---|---|---|
| Label | x | y | z | site | Occupancy | Multiplicity |
| Cu | 0 | 0 | 0 | 8a | 1.000(3) | 8 |
| Rh | 0.625 | 0.625 | 0.625 | 16d | 0.831(2) | 16 |
| Pt | 0.625 | 0.625 | 0.625 | 16d | 0.0919(2) | 16 |
| Se | 0.383146(3) | 0.383146(3) | 0.383146(3) | 32e | 1.000(3) | 32 |

| **Cu(Rh$_{0.88}$Pt$_{0.12}$)$_2$Se$_4$** | $R_{wp}$ = 5.13 %, $R_p$ = 3.69%, $R_{exp}$ = 1.11%, $\chi^2$ = 21.28 | | | | | |
|---|---|---|---|---|---|---|
| Label | x | y | z | site | Occupancy | Multiplicity |
| Cu | 0 | 0 | 0 | 8a | 1.000(3) | 8 |
| Rh | 0.625 | 0.625 | 0.625 | 16d | 0.802(2) | 16 |
| Pt | 0.625 | 0.625 | 0.625 | 16d | 0.1247(13) | 16 |
| Se | 0.3822(5) | 0.3822(5) | 0.3822(5) | 32e | 0.998(3) | 32 |

Table. S2 The proportion of each phase derived from refinement using Reference Intensity Ratio (RIR) method (RIR$_{CuRh2Se4}$ = 6.96, RIR$_{RhSe2}$ = 5.32).

| content | CuRh$_2$Se$_4$ | | | RhSe$_2$ | | |
|---|---|---|---|---|---|---|
| | peak position (°) | Integrated Intensity (cps·deg) | Ratio (%) | peak position (°) | Integrated Intensity (cps·deg) | Ratio (%) |
| $x = 0$ | 34.8782 | 19118 | 100 | - | - | - |
| $x = 0.015$ | 34.8696 | 18835 | 98.45 | 29.622 | 175 | 1.55 |
| $x = 0.02$ | 34.894 | 21267 | 100 | - | - | - |
| $x = 0.04$ | 34.8569 | 30673 | 95.4 | 29.668 | 940 | 4.6 |
| $x = 0.05$ | 34.809 | 18900 | 96.4 | 29.632 | 497 | 3.6 |
| $x = 0.06$ | 34.8075 | 11583 | 100 | - | - | - |
| $x = 0.07$ | 34.81 | 24227 | 99.754 | 29.622 | 256 | 0.246 |
| $x = 0.09$ | 34.8043 | 39088 | 100 | - | - | - |
| $x = 0.12$ | 34.7953 | 37652 | 95.84 | 29.672 | 1084 | 4.16 |

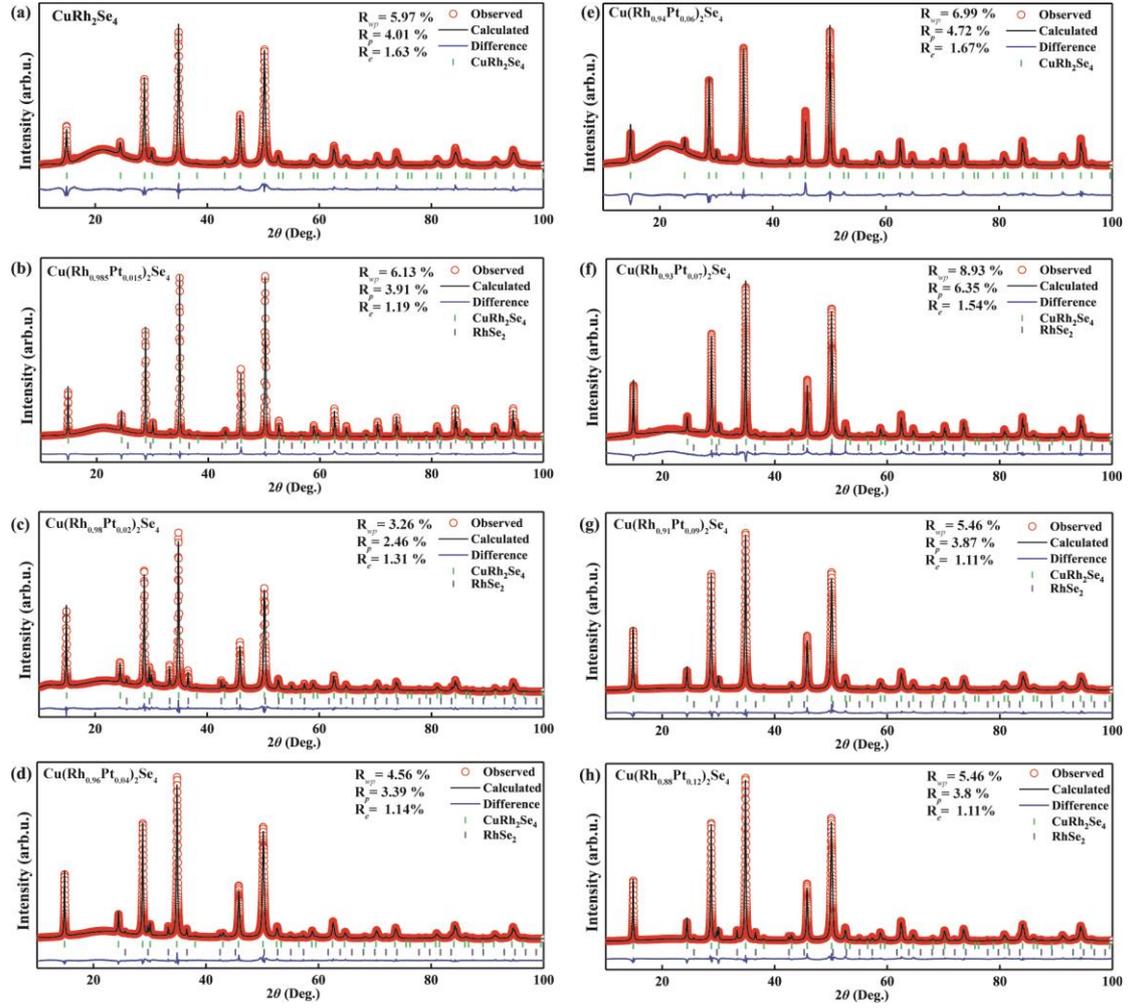

**Fig. S1** The XRD result after Rietveld refinement for (a) $CuRh_2Se_4$, (b) $Cu(Rh_{0.985}Pt_{0.015})_2Se_4$, (c) $Cu(Rh_{0.98}Pt_{0.02})_2Se_4$, (d) $Cu(Rh_{0.96}Pt_{0.04})_2Se_4$, (e) $Cu(Rh_{0.94}Pt_{0.06})_2Se_4$, (f) $Cu(Rh_{0.93}Pt_{0.07})_2Se_4$, (g) $Cu(Rh_{0.91}Pt_{0.09})_2Se_4$, (h) $Cu(Rh_{0.88}Pt_{0.12})_2Se_4$.

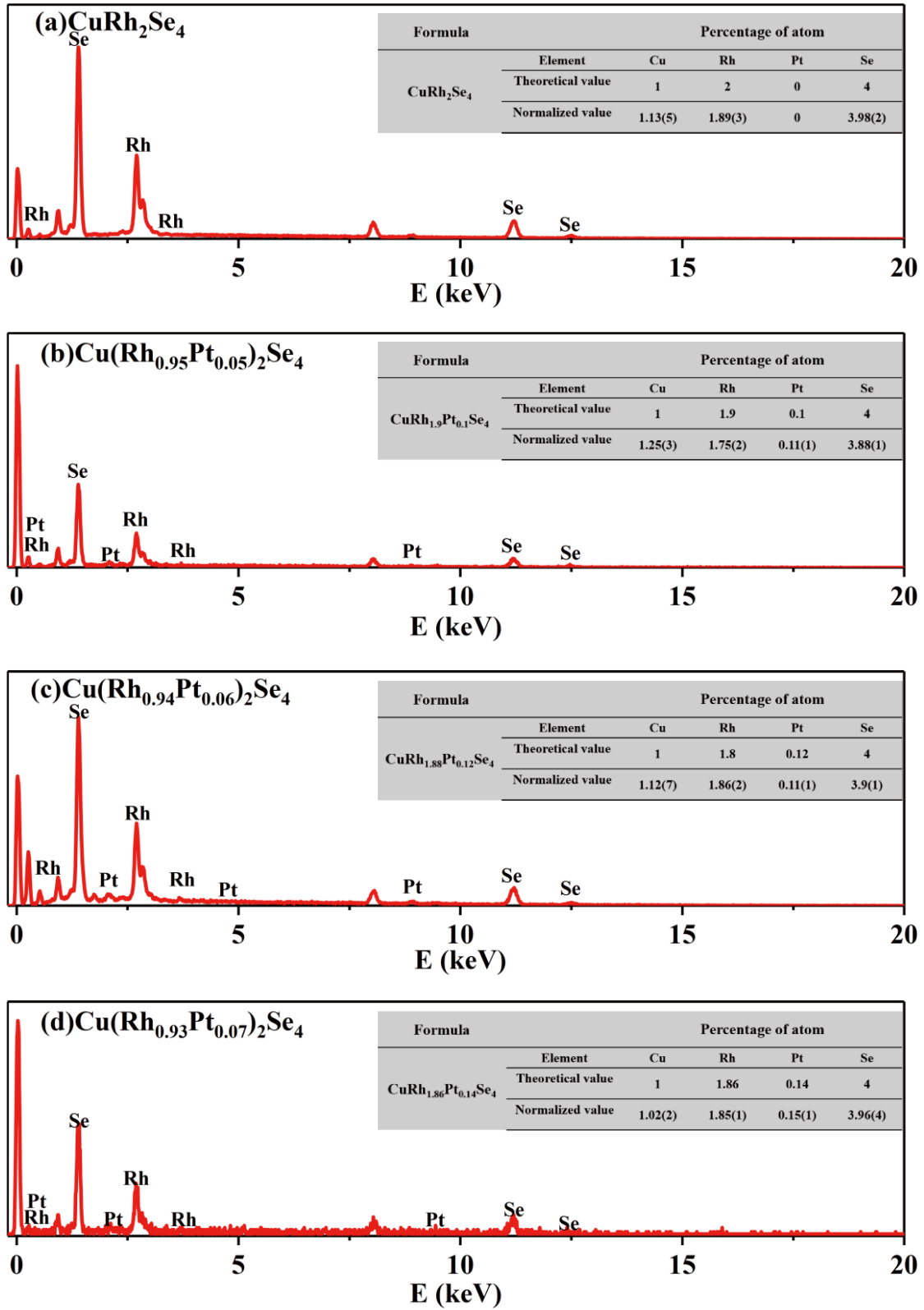

**Fig. S2** The EDXS spectrum of (a) $CuRh_2Se_4$, (b) $Cu(Rh_{0.95}Pt_{0.05})_2Se_4$, (c) $Cu(Rh_{0.94}Pt_{0.06})_2Se_4$, (d) $Cu(Rh_{0.93}Pt_{0.07})_2Se_4$ and the obtained percentage of atoms.

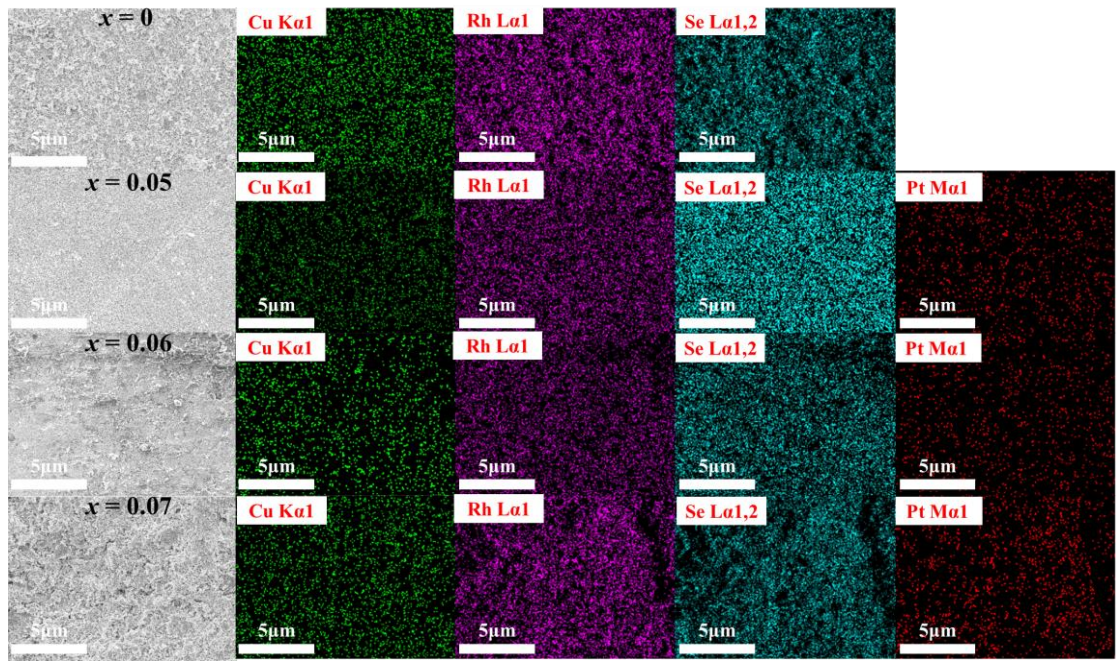

**Fig. S3** EDXS mappings of Cu(Rh$_{1-x}$Pt$_x$)$_2$Se$_4$ ($x$ = 0, 0.05, 0.06, 0.07).

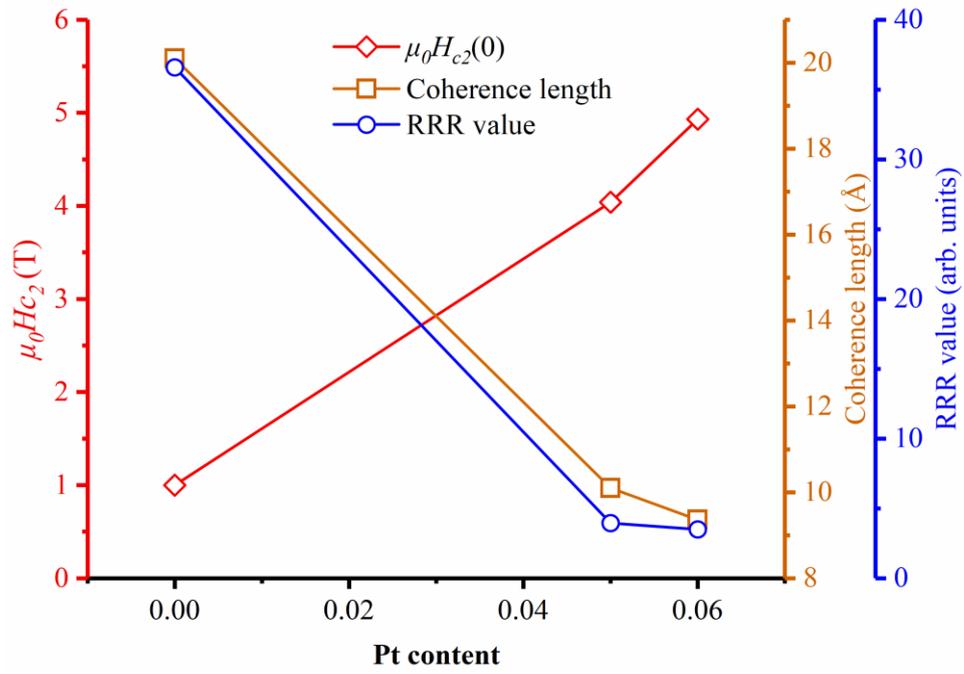

**Fig. S4** The relationship between upper critical fields ($\mu_0H_{c2}$) and coherence length ($\xi_{GL}$) and reduced residual resistivity ratio (RRR).